\documentclass{PoS}

\usepackage{amsmath}
\newcommand{\rea}{{\rm Re\,}}
\newcommand{\ima}{{\rm Im\,}}

\title{Enhanced non-quark-antiquark and non-gluebal $N_c$ behavior of light scalar mesons}

\ShortTitle{Enhanced non-quark-antiquark and non-gluebal $N_c$ behavior of light scalar mesons}

\author{\speaker{Guillermo Rios}\\%
  Departamento de F\'isica Te\'orica II, Universidad Complutense de Madrid;\\
  Departamento de F\'{\i}sica, Universidad de Murcia\\
  E-mail: \email{griosmar@fis.ucm.es}}

\author{Jenifer Nebreda\\
  Departamento de F\'isica Te\'orica II, Universidad Complutense de Madrid;\\
  Helmholtz-Institut f\"ur Strahlen- und Kernphysik (Theorie), 
  Universit\"at Bonn\\
  E-mail: \email{jenifer.nebreda@fis.ucm.es}}       

\author{Jose R. Pelaez\\
        Departamento de F\'isica Te\'orica II, Universidad Complutense de Madrid\\
        E-mail: \email{jrpelaez@fis.ucm.es}}

\abstract{We report our results on the nature of the
lightest scalar resonances, where we show that a $\bar qq$ or glueball interpretation
of the scalars $f_0(600)$ and $K^*_0(800)$ requires
a very unnatural fine tuning to satisfy $1/N_c$--expansion predictions
for $\bar qq$ or glueball states, which is not needed in the case of 
the lightest vector mesons $\rho(770)$ and $K^*(892)$. For this we
consider scattering observables whose value is fixed to 1 for $\bar qq$ and
glueball states up to corrections suppressed by more than one power of
$1/N_c$, thus enhancing contributions of other nature. 
This allows us to evaluate these observables and
check the $1/N_c$ predictions at $N_c=3$ 
without the need to extrapolate to unphysical $N_c$ values.
This is done using recent and very precise dispersive
$\pi\pi$ and $\pi K$ scattering data analyses. }

\FullConference{Sixth International Conference on Quarks and Nuclear Physics\\
                April 16-20, 2012\\
                Ecole Polytechnique, Palaiseau,  Paris}

\begin{document}

\section{Introduction}

Light scalar mesons are an object of great interest in hadron and nuclear
physics. They are largely responsible for the attractive 
part~\cite{Johnson:1955zz} of the 
nucleon--nucleon interaction; some have the quantum numbers of the
lightest glueball, which is interesting for the non--abelian nature of
Quantum Chromodynamics (QCD); also, some have the quantum numbers of the vacuum, so they
should play a relevant role in the spontaneous Chiral Symmetry breaking of QCD. 
However, the precise properties of the light scalar mesons, as their nature,
spectroscopic classification, and even their existence ---as in the case of the
$K^*_0(800)$--- are still the subject of an intense debate.
Regarding their spectroscopic nature, several 
models~\cite{Jaffe:1976ig} suggest that they might not be of
ordinary $\bar qq$ nature, but of other kind of spectroscopic classification,
such as tetraquarks, meson--meson molecules, glueballs, or a complicated mixture of all these.

A powerful tool to study the spectroscopic nature of mesons is the
QCD $1/N_c$ expansion~\cite{Witten:1980sp}. It is valid in 
the entire energy range 
and gives a clear definition of different spectroscopic components 
in terms of their mass  and width 
$1/N_c$ scaling,
which is well known for $\bar qq$ and glueball states.
By combining the $1/N_c$ expansion with Chiral Perturbation Theory
(ChPT)~\cite{chpt} unitarized with the Inverse Amplitude Method~\cite{IAM}, 
some of us studied~\cite{Pelaez:2003dy,Pelaez:2006nj} 
the $1/N_c$ behavior of light
resonances. It was found that whereas the $\rho(770)$ and $K^*(892)$
vectors behave predominantly as expected for $\bar qq$ states,
the scalars $f_0(600)$ and $K^*_0(800)$ do not~\cite{Pelaez:2003dy}. 
However, the two--loop analysis~\cite{Pelaez:2006nj} 
showed that a possible subdominant $\bar qq$ component for the $f_0(600)$ may
exist, but with a mass around 1 GeV or more. 

In~\cite{Pelaez:2003dy,Pelaez:2006nj} unitarized ChPT was used to
change $N_c$ and study the $1/N_c$ scaling of the mass and width of the 
light resonances generated.
However, the $1/N_c$ leading
$\bar qq$ scaling, $M=O(1)$, $\Gamma=O(1/N_c)$ receives subleading corrections
suppressed by $1/N_c$,
and for physical $N_c=3$ this may not seem a large suppression.
Thus, we report here our 
results~\cite{Nebreda:2011cp} using
adimensional observables with corrections suppressed further than $1/N_c$,
that allow us to obtain conclusions directly from real data at $N_c=3$,
without the need to extrapolate to larger $N_c$ using unitarized ChPT.

The observables mentioned above are related to the three different
criteria commonly used to identify resonances in elastic two--body scattering,
which are equivalent for large $N_c$. One of these criteria is the position of the
pole associated to the resonance in the unphysical sheet, $s_R$, which gives a 
definition of the resonance mass and width, $s_R=m_R^2-im_R\Gamma_R$.
A second possibility is to define the mass as the energy at which the phase shift 
reaches $\pi/2$, which for both $\pi\pi$ and $\pi K$ scalar scattering phase shifts occur 
relatively far from the pole position. Third, the resonance mass can also be
identified with the point where the phase derivative is maximum.
The relation between the first two criteria, which are equivalent up to 
$O(1/N_c^2)$ corrections for $\bar qq$ states~\cite{Nieves:2009kh}, 
was studied in~\cite{Nieves:2009kh}
for the $f_0(600)$ with a relatively inconclusive result about its assumed
$\bar qq$ nature. A more reliable parametrization and better data were called for
and we will use it here with more conclusive results.

Thus in section~\ref{observables} we define and obtain the $1/N_c$ scaling
of the observables used to test
the $1/N_c$ predictions suppressed by more than one power of $1/N_c$.
These are related to the phase shift and its derivative evaluated at the
resonance ``pole'' mass $m_R^2={\rm Re\,}(s_R)$. In section~\ref{results}
we discuss the results obtained, where we see that the coefficients needed
for considering the $f_0(600)$ and $K^*_0(800)$ as $\bar qq$ or glueball states are 
unnaturally large by two orders of magnitude.

\section{Highly suppressed $1/N_c$ observables}
\label{observables}

Consider the elastic scattering of two mesons with a resonance
associated to a pair of conjugate poles on 
the unphysical sheet of the scattering amplitude,
located at $s_R=m_R^2\pm im_R\Gamma_R$, where $m_R$ and $\Gamma_R$ are the
resonance mass and width. It was found in~\cite{Nieves:2009kh} that if 
the resonance behaves as a $\bar qq$ state, i. e., $m_R=O(1)$, $\Gamma_R=O(1/N_c)$, 
then the phase shift satisfies
\begin{equation}
  \label{delta}
  \delta(m_R^2)=\frac{\pi}{2}-
  \underbrace{\frac{\rea t^{-1}}{\sigma}\bigg|_{m_R^2}}_{O(N_c^{-1})}+O(N_c^{-3}),
  \qquad
  \delta'(m_R^2)=-\underbrace{\frac{(\rea t^{-1})'}{\sigma}\bigg|_{m_R^2}}_{O(N_c)}
  +O(N_c^{-2}),
\end{equation}
where $t(s)$ is the scattering partial wave, $\sigma=2k/\sqrt{s}$, $k$ is the
center of mass momentum of one of the mesons and $s$ is the usual Mandelstam variable.
The prime denotes derivatives with respect to $s$. Note that the 
subleading $1/N_c$ corrections are suppressed by two powers of $1/N_c$.
This particular $1/N_c$ counting, as shown in~\cite{Nieves:2009kh}, 
comes from the expansion of the real and imaginary parts of the pole equation,
as we detail next.

The inverse of the partial wave,
which generically scales as $N_c$, can be written as $t^{-1}=R+iI$,
where $R$ and $I$ are
analytic functions that coincide with the real and imaginary parts of $t^{-1}$
over the right cut, i. e., $R(s)=\rea t^{-1}(s)$ and $I(s)=\ima t^{-1}(s)=-\sigma(s)$
for $s>s_{th}$. Then, the inverse partial wave on the second sheet
is given by $t^{-1}_{II}=R-iI$, and the equation for the resonance 
pole position, $t^{-1}_{II}(s_R)=0$, can be written as $R(s_R)=iI(s_R)$.
If the resonance is a $\bar qq$ state, $m_R=O(1)$
and $\Gamma_R=O(N_c^{-1})$, and we take the real and 
imaginary parts of the expansion of the pole equation 
around $m_R^2$, we arrive at
\begin{equation}
  \begin{aligned}
    \rea t^{-1}(m_R^2)&=\underbrace{m_R\Gamma_R\left[\frac{m_R\Gamma_R}{2}
        (\rea t^{-1})''_{s=m_R^2}
        -\sigma'(m_R^2)\right]}_{O(N_c^{-1})}+O(N_c^{-3}), \quad \label{retinvexpansion}\\
    (\rea t^{-1})'_{s=m_R^2}&=
    \underbrace{\frac{\sigma(m_R^2)}{m_R\Gamma_R}}_{O(N_c)}+O(N_c^{-1}).
  \end{aligned}
\end{equation}
Since the
expansion parameter $im_R\Gamma_R\sim 1/N_c$ is purely imaginary,
the different orders in the expansion, which are suppressed by the
corresponding $1/N_c$ factors, are
real or purely imaginary alternatively. Then, 
when taking the real and imaginary parts of the equation, the 
different orders are suppressed by two powers of $1/N_c$, as
shown in Eqs.~\eqref{retinvexpansion}, from where we also see 
that the inverse amplitude scales as $1/N_c$ instead of as the generic $N_c$
when evaluated at $m_R^2$. Then,
Eqs.~\eqref{delta} are obtained noting that
the phase shift $\delta(s)$ satisfies $\delta-\pi/2=-\arctan(\rea t^{-1}/\sigma)$,
and using Eqs.~\eqref{retinvexpansion} to expand the
$\arctan$ function in $1/N_c$ powers.

We can now define from Eqs.~\eqref{delta}
the following adimensional observables,
\begin{equation}
  \label{adef}
  \frac{\frac{\pi}{2}- \rea t^{-1}/\sigma}{\delta}\Big\vert_{m_R^2}
  \equiv\Delta_1=1+\frac{a}{N_c^3},
  \qquad
  -\frac{[\rea t^{-1}]'}{\delta'\sigma}\Big\vert_{m_R^2}\equiv\Delta_2=1
  +\frac{b}{N_c^2},
\end{equation}
whose value should be one
for predominantly $\bar qq$ resonances up to $O(1/N_c^3)$ and
$O(1/N_c^2)$ corrections, respectively. We have written explicitly the 
corresponding $1/N_c$ powers in the subleading terms, so the
coefficients $a$ and $b$ should naturally be $O(1)$ or less.
Note that it is relatively simple to make $a$ and $b$ much smaller
than one by taking into account higher order contributions of
natural size, but very unnatural to make them much larger.
In the case of a glueball nature of the resonance, whose mass and 
width scale as $m_R=O(1)$ and $\Gamma_R=O(1/N_c^2)$, the above 
derivations can be repeated, but now the subleading corrections are
even more suppressed since the width scales as $1/N_c^2$ instead of only
$1/N_c$. Then, for a glueball resonance, the observables $\Delta_1$ and
$\Delta_2$ satisfy
\begin{equation}
  \label{a_b_glueball}
  \Delta_1=1+\frac{a'}{N_c^6},\qquad \Delta_2=1+\frac{b'}{N_c^4},
\end{equation}
where $a'$ and $b'$ should of natural $O(1)$ size.

In the following section we will calculate these 
observables to see how well
the above predictions for $\Delta_1$ and $\Delta_2$ are fulfilled 
assuming a $\bar qq$ nature (or also glueball for the $f_0(600)$) for the
resonances found in elastic of $\pi\pi$ and $\pi K$ scattering.

\section{Results}
\label{results}

In Table~\ref{tab:Nc3results} we show the values of the
$a$ and $b$ parameters for the lightest resonances found in $\pi\pi$ and
$\pi K$ scattering, which have been calculated from the 
data analyses that we detail below. Let us first note that
{\it for the $\rho(770)$ and $K^*(892)$ vector resonances all parameters
are of order one or less, as expected for $\bar qq$ states}.   
In contrast, {\it for the $f_0(600)$ and $K^*_0(800)$ scalar resonances we 
find that all parameters are larger, by two orders of magnitude,
than expected for $\bar qq$ states}. 
This is one of our main results and make 
the $\bar qq$ interpretation of both scalars extremely unnatural. 
\begin{table}[b]
  \centering
  \begin{tabular}{crrrr}
    & $\rho(770)$ & $K^*(892)$ & $f_0(600)$ & $K^*_0(800)$ \\\hline
\rule[-1mm]{0mm}{5mm}$a$    &    $-0.06\pm0.01$  &  0.02  &  $-252^{+119}_{-156}$ & -2527 \\
\rule[-1mm]{0mm}{5mm}$b$    &  $0.37 ^{+0.04}_{-0.05}$    & 0.16 &$77^{+28}_{-24}$ & 162 \\
\hline
  \end{tabular}
  \caption{Normalized coefficients of the $1/N_c$ expansion for different resonances.
For $\bar qq$ resonances, all them are expected to be of order one or less.}
  \label{tab:Nc3results}
\end{table}

The data analyses that we have used in each case are the following.
For the $\pi\pi$ scattering phase shifts we use the very precise
and reliable output of the data analysis in \cite{GarciaMartin:2011cn}
constrained to satisfy Roy equations, once subtracted
Roy--like equations (GKPY equations) and forward dispersion relations, which is 
therefore model independent and specially suited to obtain the $f_0(600)$ pole
\cite{nosotros}. This analysis is also in good agreement with previous
dispersive result based on Roy equations~\cite{CGL}. 
For the case of isospin $1/2$ scalar channel of $\pi K$ scattering,
where we find the $K^*_0(800)$, 
we have also used the rigorous
dispersive calculation in~\cite{DescotesGenon:2006uk} that uses
Roy-Steiner equations, although in this case we can only provide a central value.
For the isospin $1/2$ vector channel of $\pi K$ scattering, where
we find the $K^*(892)$, there are
no very precise purely dispersive descriptions of data, so we use unitarized
ChPT in the form of the elastic IAM~\cite{IAM}.
We have checked that using the IAM for the $\rho(770)$ we obtain results
within 50\% of the results using the GKPY dispersive representation.
Since the $K^*(892)$ is narrower than the $\rho(770)$, the IAM is likely
to provide a better approximation than in the $\rho(770)$ case, but even with
that 50\% uncertainty we can check that the $a$ and $b$ parameters are smaller 
than one.

One might argue that, since the first of Eqs.~\eqref{delta} comes from the
expansion of $\arctan(x)=x-x^3/3+...$, the correction $a/N_c^3$ to $\Delta_1=1$
is really the cube of a $O(N_c^{-1})$ quantity, $(\tilde a/N_c)^3$, where now
the coefficient that should be natural is $\tilde a$ instead of $a$.
That explains the very small values obtained for the $\rho(770)$ and the
$K^*(892)$, that come from $a=\tilde a^3/3$, with $\tilde a=0.56\pm0.03$ and
$\tilde a=-0.4$ for the $\rho(770)$ and $K^*(892)$ respectively, which are quite
natural values. For the $f_0(600)$ and the $K^*_0(800)$ we obtain
$\tilde a=9.1^{+1.3}_{-2.5}$ and $-19.6$, still rather unnatural values.
In the case of $\Delta_2$, where the corrections are only suppressed
by $1/N_c^2$ instead of $1/N_c^3$, we do not find this issue, because
the $b/N_c^2$ term is not the square of a natural $1/N_c$ quantity,
\begin{equation}
\frac{b}{N_c^2}=\frac{\rea t^{-1}}{\sigma}
\Big[\frac{\sigma'}{(\rea t^{-1})'}-\frac{\rea t^{-1}}{\sigma}\Big]+O(N_c^{-4}).
\end{equation}
Despite containing a cancellation between two $1/N_c$ terms, its value for the 
$\rho(770)$ and $K^*(892)$ is rather natural. However, the value for the scalars 
is almost two orders of magnitude larger than expected for predominantly $\bar qq$
states.

In the case of a glueball interpretation of the $f_0(600)$, the coefficients
$a'$ and $b'$ from Eqs.~\eqref{a_b_glueball} are even more unnatural,
this time too large by three or four orders of magnitude,
$a'=-6800^{+3200}_{-4200}$ and $b'=2080^{+760}_{-650}$. In other words, 
a very dominant or pure glueball nature for the $f_0(600)$ is
very disfavored by the $1/N_c$ expansion.
Of course, as in the $\bar qq$ case we could worry about the fact that, due to
the $\arctan(x)=x-x^3/3+...$ expansion, the $a'$ should be interpreted as
$a'=\tilde a'/3$. However, even with that interpretation we would still
find $\tilde a'=27^{+5}_{-7}$, again rather unnatural.
Once more, in the case of $b'$, its value is genuinely unnatural, 
disfavoring the glueball interpretation.

Finally, in~\cite{Nebreda:2011cp} we also showed that what really happens for
the scalars is that they do not even follow the $1/N_c$ expansion of $\bar qq$
or glueball states given in Eqs.~\eqref{adef} and \eqref{a_b_glueball}.
This was done by calculating the $1/N_c$ scaling of the quantities $\Delta_i-1$ for the
different resonances using the Inverse Amplitude Method, 
where the $1/N_c$ expansion can be implemented through 
the ChPT low energy constants. We refer however the reader to
our original work~\cite{Nebreda:2011cp} for further details.

\section{Summary}

We have reviewed our results in~\cite{Nebreda:2011cp} where we study the
$1/N_c$ expansion of elastic meson--meson scattering phase shifts around
the pole mass of a $\bar qq$ or glueball resonance. In particular, we
have defined the observables~\eqref{adef} and \eqref{a_b_glueball},
whose value is fixed to one up to corrections suppressed by more than
one power of $1/N_c$ for $\bar qq$ or glueball states. Using
very precise dispersive analyses of 
$\pi\pi$ and $\pi K$ scattering data we have shown that a $\bar qq$ 
or glueball interpretation for the $f_0(600)$ or $K^*_0(800)$ 
needs unnaturally large coefficients in the expansion.
Thus, a predominant $\bar qq$ or glueball nature for these resonances
is heavily disfavored by the $1/N_c$ expansion, and
this has been shown without extrapolating beyond $N_c=3$.
However, when extrapolating to larger $N_c$ using the IAM, we checked 
in~\cite{Nebreda:2011cp} that the scalars do not follow the pattern
of the $1/N_c$ expansion expected for $\bar qq$ or glueball states.

This work is partially supported by the FPA2011-27853-C02-02 Spanish grant.

\end{document}